\documentclass[aps,prl,showpacs,reprint]{revtex4-1}
\usepackage{graphicx}

\usepackage{epsfig}
\usepackage[]{natbib}
\usepackage{xspace}
\usepackage{bbold}
\usepackage{tabu}

\usepackage{amsmath}
\usepackage{amssymb}
\usepackage{mathrsfs}
\usepackage{mathtools}
\usepackage{color}
\usepackage{array}

\makeatletter
\g@addto@macro\bfseries{\boldmath}
\makeatother

\begin{document}
 
  \title{
Transitions in 
the ergodicity of subrecoil-laser-cooled gases
}

\author{Eli Barkai}
  \affiliation{Department of Physics, Institute of Nanotechnology and Advanced Materials, Bar-Ilan University, Ramat-Gan 52900,
  Israel}
\author{G\"unter Radons}
\affiliation{
Institute of Physics, Chemnitz University of Technology, 09107 Chemnitz, Germany}
\affiliation{
Institute of Mechatronics, 09126 Chemnitz, Germany. }
  \author{Takuma Akimoto}
\affiliation{
Department of Physics, Tokyo University of Science, Noda, Chiba 278-8510, Japan}

  \date{\today}

  \begin{abstract}

With 
 subrecoil-laser-cooled atoms one may reach
 nano-Kelvin temperatures while the ergodic properties
of these systems do not follow usual statistical laws. Instead, due to
an ingenious trapping mechanism in momentum space, 
power-law-distributed 
sojourn  times are found  for the cooled  particles.  
Here, we show how this gives rise to 
a statistical-mechanical framework based on infinite ergodic theory,
which replaces ordinary ergodic  statistical physics of a thermal gas of atoms.
In particular, the energy of the system exhibits a sharp 
discontinuous 
 transition
in its ergodic properties. Physically this  is controlled by the
fluorescence rate, but more profoundly it
 is a manifestation of a transition for 
any observable, from  being  an integrable to becoming a
 non-integrable observable,
 with respect
to the infinite (non-normalised)  invariant density.

  \end{abstract}
\maketitle

Laser cooled atoms are important for fundamental  and practical applications. 
It is well known that L\'evy statistics describes the unusual  properties of
the
cooling processes \cite{Bardou,CT,Zoller,Sagi,Kessler1,Dechant,AghionPRX,Renzoni}. 
For subrecoil laser cooling a special
 atomic
trap in momentum space is engineered.  The most efficient cooling is found when
a mean trapping time, defined more precisely below, diverges \cite{CT}. 
The fact that the characteristic time diverges implies that 
the ergodic properties of these systems must differ
from those of standard gases \cite{CT,WEB,Saubamea}. 
Ergodicity is a fundamental aspect of statistical mechanics 
that implies that the time and ensemble averages coincide. This is found when the measurement time is made long compared to the time scale of the dynamics. 
However, in the context of subrecoil  laser cooling  this time 
diverges, and hence no matter how long one measures, deviations from standard
ergodic theory are prominent. Given that lasers replace heat baths in many cooling experiments, what are the ergodic properties of the system?
In other words, what  replaces the usual ergodic statistical framework?
Our goal is to show how tools
of infinite ergodic theory describe the statistical properties
of the ensemble and corresponding  time averages of the 
subrecoil-laser-cooled atoms.

 Infinite ergodic theory was investigated by mathematicians \cite{Darling,Aaronson,Zweim}
 and more recently 
in Physics \cite{PRLKorabel,Kessler,Miya,Akimoto2012,Holz,Shinkai,Kantz,Burioni,Erez,Erez1,Sato,Celia,PRETakuma,Artuso}. 
In generality, infinite ergodic theory
 deals with a peculiar non-normalised density, describing the long time limit
of a system,
called below the infinite invariant density. Previous works
 in the field of subrecoil laser cooling \cite{CT,Bertin}
foresaw this quasi-steady state.
 We will see  
 how to  use this tool to investigate the ensemble averages of
physical observables.  
However,  this does not give direct information on the time averages
and here we will develop physical and mathematical
insights on the latter. The basic question is how to relate ensemble
and time averages, even when ergodicity in its standard formulation is broken.  
In particular we investigate the energy of the system. Since the
atoms are non-interacting, in a classical thermal setting the energy of
the atoms per degree of freedom would be $k_B T/2$,  as a consequence 
of Maxwell's velocity distribution.
At variance with this, we will show, that the energy of a 
subrecoil-laser-cooled 
gas is obtained under certain conditions with a non-normalisable
invariant density.  A sharp
 transition is exposed, in the statistical
properties of the energy, when the fluorescence rate, given by
 $R(v)\propto |v|^{\alpha}$ in
the vicinity of zero velocity, is varied, 
more precisely when $\alpha=3$.  Since 
experimental work demonstrates the capability of a variation of
$\alpha$ from
$\alpha=2$ to $\alpha=4$ or, in principle, $\alpha=6$ etc, 
the rich phase diagram of ergodic properties which we find seems to be within 
reach of experimental investigations.  This new type of  transition
is related
to the fact that the energy observable can switch from being an integrable
observable, with respect to the infinite density, to being non-integrable.

Let $\rho(v,t)$  be the probability density function (PDF)
 of the speed $v>0$ of the atoms at time $t$. A master
equation governs its  evolution 
with typical gain and loss terms
\begin{equation}
\frac{\partial \rho \left( v,t\right) }{\partial t}=\int_{0}^{\infty }\left[
W(v^{\prime }\rightarrow v)\rho \left( v^{\prime },t\right) -W(v\rightarrow
v^{\prime })\rho \left( v,t\right) \right] {\rm d}v^{\prime }.
\label{Master}
\end{equation}
The transition rate from $v$ to $v'$ is $W(v \rightarrow v')= R(v) f(v')$.
Here $R(v)=1/\tau(v)$ is the fluorescence rate and $f(v)$ denotes the speed PDF after the atom experiences a spontaneous jolt. The specific forms of these functions, investigated previously \cite{CT}, are crucial for our analysis, but now we treat the problem in generality. It is natural to consider the long
time limit of the PDF $\rho(v,t)$. For this aim, as usual, we consider the steady-state condition $\partial {\rho}(v,t) / \partial t=0$. This gives
the time-independent 
solution $\rho^*(v)= \tau(v) f(v)/Z$, where $Z$, if it exists,
is a time independent
constant obtained from the normalization condition. 
We treat the opposite non-normalizable case which is highly relevant.

Indeed, for subrecoil laser cooling, once the atom is close to a halt,
the rate of change of speed is becoming small, and that drives the system
to low temperatures, e.g. nano-Kelvins. More specifically,
when the fluorescence time  $\tau(v)$ is given by
  $\tau(v)\sim c v^{-\alpha}$ for 
$v\rightarrow 0$,   then $\rho^{*}(v)$ becomes 
non-normalizable, for $\alpha>1$.
 Importantly, as we will shortly explain, the non-normalizable function $\rho^{*}(v)$ is not an object we should ignore. An analysis of the master 
equation
(see SM) and following the footsteps of \cite{CT,Bertin} yields
\begin{equation}
\lim_{t \to \infty} Z(t) \rho(v,t) = \tau(v) f(v) \equiv {\cal I}(v),
\ \mbox{for} \ \ \alpha>1.
\label{eq02}
\end{equation}
A calculation gives  $Z(t) = \pi \Gamma(1 + \gamma) f(0) c^\gamma t^{1-\gamma}/\sin \pi \gamma$ with $\gamma=1/\alpha$. In contrast, if $\alpha<1$ 
the usual normalization $Z= \int_0 ^\infty f(v) \tau(v) {\rm d} v$, is found. For the case under study, $\gamma<1$ or  $\alpha>1$, the integration over ${\cal I}(v)$ diverges, due to the  small $v$ behavior of $\tau(v)$, and hence ${\cal I}(v)$ is called 
an infinite invariant density. Note that in this case,
on the left-hand side of  Eq. (\ref{eq02}),  we multiply the normalized density 
$\rho(v,t)$
with a function $Z(t)$ that is increasing with time, therefore  it is not surprising that on the right-hand side we find in the long-time limit
 a non-normalizable function. The question we tackle is what is the
physical meaning of this state? And how do we use the infinite invariant density to construct a non-trivial ergodic theory for the gas?   

One can argue that
the infinite density, as a stand alone, cannot be the full picture, as
it is non-normalizable. Indeed, for long but finite times,
the density of the velocity of the
 particles can be described by two regimes.  As the density is evolving, it is shrinking in its width peaking on zero speeds.
In the inner region, the density $\rho(v,t)$ exhibits a scaling solution, given by
\cite{CT}  
$\rho(v,t) \sim t^\gamma g(v t^\gamma)$ with 
\begin{equation}
g(x) = { {\cal N} \over \gamma x} \exp\left( - {x^{1/\gamma} \over c}  \right)\int_0 ^x \exp\left( {z^{1/\gamma} \over c}  \right) {\rm d} z. 
\label{eq03}
\end{equation}
This describes velocities of order 
$1/t^\gamma$, i.e. an inner region of the packet, which 
for $t\to \infty$ goes to zero (see SM). 

The scaling solution Eq. (\ref{eq03})
 and the infinite density Eq. (\ref{eq02})  are not separable and
both together yield a complete description of the packet.
Mathematically, the two solutions match at intermediate velocities (see
SM). 
Eq. (\ref{eq03})
 predicts that the full width at half maximum of the velocity PDF decays like
$t^{-1/2}$ for $\alpha=2$  and as $t^{-1/4}$ for $\alpha=4$.
These theoretical predictions
 were indeed observed in the laboratory  with Cesium \cite{CT1,Reichel},
see also quantum  Monte Carlo simulation and experiments
with Helium
in \cite{CT,Saubamea}. However, the scaling solution Eq. (\ref{eq03})
 exhibits what
might seem naively as an unphysical feature. If we consider the realistic case $\alpha=2$, we find that the scaling function as a stand alone predicts that the
second moment of the velocity, namely the kinetic energy, is infinite,
and in this sense the system cannot be considered cold. 
This is due to the fat tail of the scaling function for large $v$.
As explained below, this issue is cured, using the non-normalized solution
Eq. (\ref{eq02}).
In general, we need to classify observables based on 
whether they are integrable or non-integrable with respect to 
${\cal I}(v)$. These two classes of  observables have vastly different ergodic
properties, which are now investigated. 

{\em Ensemble and time averages.}
Let $v(t)$ be the random speed process of an atom. 
We now consider 
generic observables ${\cal O}[v(t)]$, and we will
study  their  time and ensemble averages. 
As examples consider
 ${\cal O}[v(t)]= v^2(t)=E_k(t)$, which is the kinetic energy
when we set $m/2=1$ 
and the indicator function ${\cal O}[v(t)] = I(v_a<v(t)<v_b)$ where 
this function equals one if the condition $v_a<v(t)<v_b$ is true,
 otherwise it is zero.
By definition the ensemble average is
$\langle {\cal O}(t) \rangle = \int_0 ^\infty {\cal O}(v) \rho(v,t) {\rm d} v$,
 so using Eq.
(\ref{eq02}),  we find in the long-time limit
\begin{equation}
\lim_{t \to \infty} Z(t) \langle {\cal O}(t) \rangle = \int_0 ^\infty {\cal I}(v) {\cal O}(v) {\rm d} v < \infty. 
\label{eq04}
\end{equation}
This equation shows that the non-normalized density ${\cal I}(v)$ is used
in the computation of ensemble averages, in a way reminiscent of the
standard averaging performed for equilibrated systems with normalized densities.
The only
condition we have is that the observable is integrable with respect
to ${\cal I}(v)$, namely that the integral exists. Since
${\cal I}(v) \sim v^{-\alpha}$ for small velocity, the kinetic energy
is an integrable observable for $\alpha<3$. However it is  non-integrable
otherwise. The critical value $\alpha=3$ or $\gamma=1/3$
 will mark an ergodicity transition
for this observable. In contrast, the indicator function is always an integrable observable, provided that $v_a>0$.

A  goal of ergodic theories is to relate the ensemble and the time
averages, denoted with an over-line, 
$\overline{{\cal O}}(t) =  \int_0 ^t {\cal O} [v(t')] {\rm d} t'/t$.
According to the standard
ergodic hypothesis $\overline{{\cal O}}(t)/\langle {\cal O} \rangle \to 1$ in the limit of long-measurement-times. In our case, the time averages remain random, and we will soon investigate their fluctuations.
To start, we consider an ensemble of paths and average over time and
then over the ensemble 
\begin{equation}
\langle {\overline {\cal O}}(t) \rangle = \left\langle {1 \over t}
 \int_0 ^t {\cal O}[v(t')] {\rm d} t'  \right\rangle = 
{1 \over t}
 \int_0 ^t \int_0 ^\infty {\cal O}(v) \rho(v,t') {\rm d} v {\rm d} t'.
\label{eqave05}
\end{equation}
Considering the
long-time limit and using Eq. 
(\ref{eq02})
\begin{equation}
\langle {\overline {\cal O}} (t) \rangle \sim
 {1  \over t} \int_0 ^t  {\rm d} t' { \int_0 ^\infty {\cal O}(v) {\cal I} (v) {\rm d} v \over  Z(t') } = { \int_0 ^\infty {\cal O}(v) {\cal I} (v) {\rm d} v \over \gamma Z(t)} 
\label{eqave06}
\end{equation}
where we used $0<\gamma<1$. This resembles the standard calculation of the
time averages, performed when the invariant density is normalizable. 
 We conclude that
for integrable observables and if $\gamma<1$
\begin{equation}
\lim_{t \to \infty} { \langle \overline {\cal O}(t) \rangle  \over  \langle {\cal O}(t) \rangle} ={ 1 \over \gamma} .
\label{eqave07}
\end{equation}
Thus we established a relation between the time average and the 
ensemble average. The latter is obtained using the infinite density ${\cal I}(v)$
 and thus this invariant
density is not merely a tool for the calculation of the ensemble average,
but rather it gives also information on the time average.  We see that
when $\gamma\to 1$ we approach standard ergodic behavior. Physically
this is related to the observation made below,
that  the mean time between velocity updates 
diverges for $\gamma<1$.

\begin{figure}
\begin{center}
\epsfig{figure=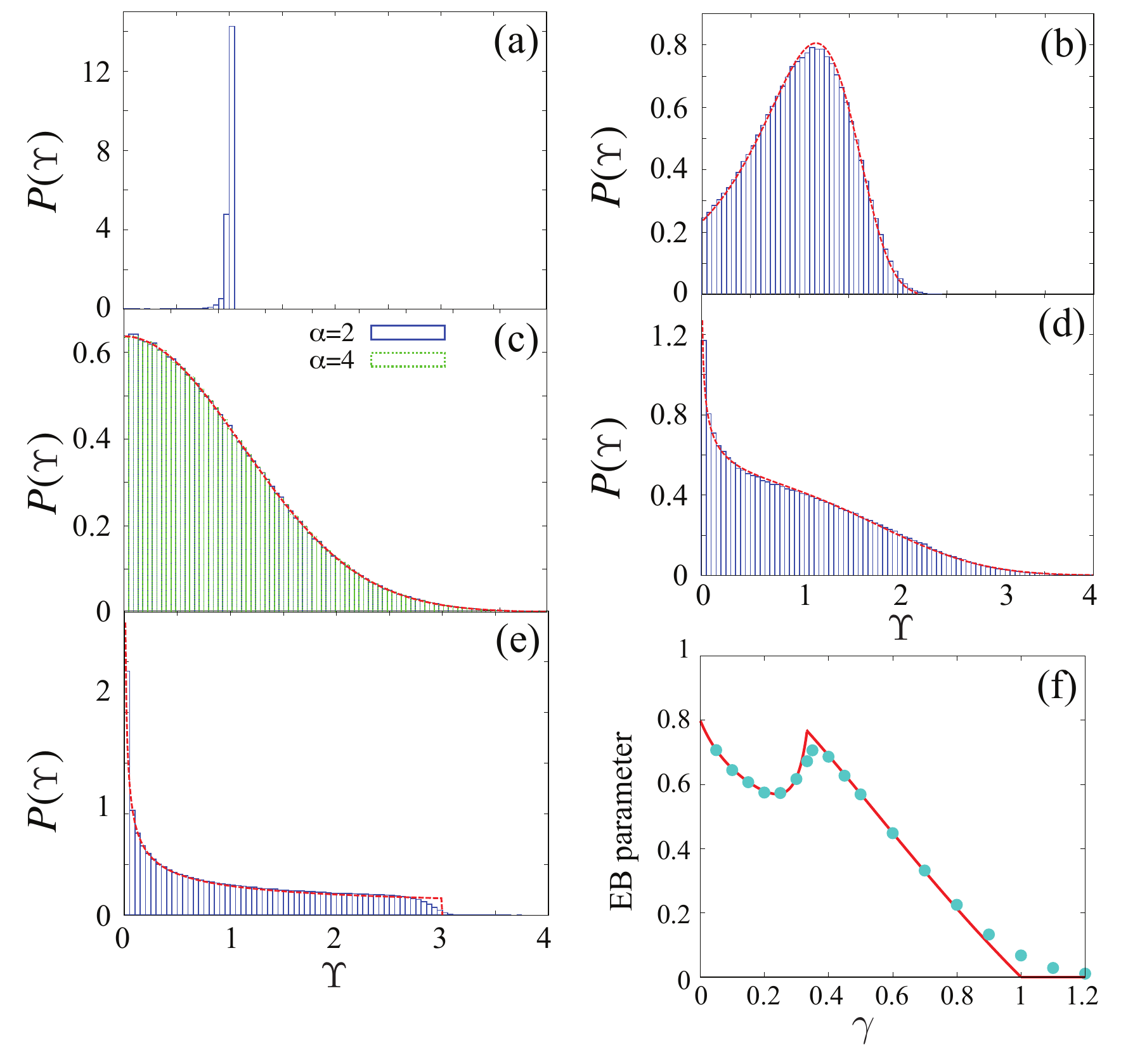, totalheight=0.42\textheight, width=0.48\textwidth,trim=0mm 0mm 0mm 0mm, clip}
\end{center}
\caption{
In the process of subrecoil laser cooling, 
the PDF $P(\Upsilon)$
 of the time averaged energy 
  $\Upsilon= \overline{E_k(t)}/\langle \overline{E_k (t)} \rangle$, exhibits a
wide range of 
physical behaviors.   
When $\gamma=1/\alpha>1$ ergodicity holds and the PDF 
 approaches a delta function in the long-measurement-time limit, see
finite time simulations in
sub-plot ${\bf a}$ with $\alpha=0.8$ (details in SM). 
When $\alpha=1.25$, the energy
is integrable with respect to the infinite measure, and $P(\Upsilon)$
is non-trivial though it has a peak close to unity, 
since at this stage  we are not too far from
the ergodic phase (plot ${\bf b}$). For $\alpha=2$ and $\alpha=4$
 we find that the fluctuations of the time average, 
are described by a half-Gaussian distribution,  presented in ${\bf c}$.
Here the
energy observable is integrable in the first case
 while it is not in the second, further  note that
the  Mittag-Leffler distribution is half-Gaussian
when $\alpha=2$ since  $\gamma=1/2$. 
When
$\alpha=6$ the energy  observable is non-integrable
and $P(\Upsilon)$  diverges for  $\Upsilon \to 0$,
indicating very large deviations from usual ergodic behavior (plot ${\bf d})$.
In the limit $\alpha \rightarrow \infty$ 
we get a discontinuous behavior, a sharp
cutoff at $\Upsilon=3$ (see plot ${\bf e}$, and simulation with
$\alpha=50$) . Finally,
the EB parameter, subplot ${\bf f}$,  exhibits a cusp at $\gamma=1/\alpha=1/3$. This value marks a transition for the  ergodic behavior  of the system,
from an energy which is  integrable with respect to the infinite measure
to non-integrable. 
In the figure simulations match theoretical predictions without fitting. 
}
\label{figsix}
\end{figure}

The time averages are functionals of the stochastic path $v(t)$
 and we  now develop
a machinery to explore their statistical properties. Here we will focus on
the kinetic  energy observable,
 due to its physical importance.
To analyse the latter we recall the stochastic process under study.
 Initially
we draw the speed $v$ from $f(v)$, the particle then remains in this state for a random time denoted $\tilde{\tau}$. This  time $\tilde{\tau}$
 is exponentially distributed with the life time $\tau(v)$
 that depends
on the velocity. The process is then renewed, namely after
waiting for a time $\tilde{\tau}$ we draw a new velocity from $f(v)$ and hence
a new life time etc. In simulations below, following \cite{CT}, we use
a uniform PDF $f(v) =1/v_{{\rm max}}$ for $v<v_{{\rm max}}$   and $\tau(v)= c v^{-\alpha}$. 
Using Eqs. (\ref{eq04},
\ref{eqave07})
\begin{equation}
\langle \overline{E}_k (t) \rangle  
\sim {\int_0 ^{v_{{\rm max}} } v^2 {\cal I} (v) {\rm d} v \over \gamma  t^{1-\gamma}}
= { \sin( \pi \gamma) \over \Gamma(1 + \gamma) \pi} { (v_{{\rm max}} )^{3-1/\gamma} \over 3 \gamma  - 1} \left( { c \over t} \right)^{1-\gamma}
\label{eqMEAN}
\end{equation}
where $\overline{E}_k(t)= \int_0 ^t E_k(t) {\rm d} t / t$. 
 This holds under two conditions, namely
 that the energy is an integrable observable, so
$1/3<\gamma$, and $\gamma<1$, otherwise we are in the regime of
standard ergodic theory.  
Interestingly,  when we take $\gamma\rightarrow 1/3$
the prefactor in
Eq. (\ref{eqMEAN}) diverges and  becomes $v_{{\rm max}}$ independent. 
This marks the entry into the
phase where the energy is non-integrable. 
After the calculation of the expectation of the time
 averaged  energy with the infinite
density ${\cal I}(v)$,  
the real challenge is to obtain the distribution of $\overline{E}_k$. 

 We rewrite
the time average
 $\overline{E}_k (t)= {\cal S}(t) /t$, where ${\cal S}(t)$ is the
action. The key idea is to investigate  the distribution of the action,
and with this to infer the ergodic properties of the process. 
We perform this task with a new form of  coupled continuous-time random
walks, which in turn is a variation of the well known  L\'evy walk
\cite{KBS,MetzKlaf,Miya1JSP,Kutner,Denisov,Albers}.   
 In the time interval of observation $(0,t)$ we have $N(t)$ random renewal
events, and the pairs of waiting times and velocities are labeled
$(v_i, \tilde{\tau}_i)$, where  $i=1, \cdots, N(t)$ ($v_1$ is the initial condition).  
We rewrite the action  ${\cal S}(t) = \sum_{i=1} ^{N(t)}  s_i + s_B(t)$ 
with $s_i=(v_i)^2 \tilde{\tau_i}$ which is reminiscent of a biased random walk process.
 The increments $s_i>0$ are constrained since the measurement time is
 $t = \sum_{i=1}^{N(t)} \tilde{\tau}_i + t_B(t)$. Here $t_B$ is the so called
backward recurrence time \cite{Godreche2001,Longest}, the time elapsing between the last update in
the process and the measurement time $t$. Similarly
 $s_B(t)=(v_{N(t)+1})^2 t_B(t)$  is the contribution to the action, from the last time interval in the sequence.

To advance the theory we need the joint PDF of action increments
 $s$ and waiting times  $\tilde{\tau}$ denoted $\phi(s,\tilde{\tau})$. This is obtained from
$\int_{0} ^{v_{{\rm max}}} {\rm d} v f(v)  \delta (s - v^2 \tilde{\tau})\exp( -\tilde{\tau}/\tau(v))/\tau(v)$ 
  which gives
\begin{equation}
\phi (s, \tilde{\tau}) = { 1 \over 2 v_{{\rm max}} \sqrt{ s \tilde{\tau} }} R \left(  \sqrt{ { s \over \tilde{\tau} }} \right) 
\exp\left[ - \tilde{\tau}  R \left(  \sqrt{ { s \over \tilde{\tau} }} \right) \right]
\label{ctrw05} 
\end{equation} 
when $0\le s \le  v_{{\rm max}} ^2\tilde{\tau}$, and zero  otherwise.
Here the waiting times $\tilde{\tau}$ and
action increments $s$ are
clearly correlated. 
Integrating over $s$ we find the PDF of the waiting times
$\psi(\tilde{\tau})$ whose fat tail reads  $\psi(\tilde{\tau}) \sim \mbox{const}
\tilde{\tau}^{- 1 - \gamma}$ with $\mbox{const} =   
c^{\gamma} \gamma \Gamma(1 + \gamma) / v_{{\rm max}}$. As mentioned,
the divergence of the mean waiting time found for $0<\gamma<1$
signals special ergodic properties
 \cite{WEB}.  

 Let $P({\cal S},t)$ be the PDF of the action at time $t$. The first step in
the analysis is to relate this PDF to Eq. (\ref{ctrw05}). Using techniques borrowed from random walk theory,  employing the renewal property of the process
and the convolution theorem, we establish this relation using Laplace transforms.
Let $\hat{P}(u,p)= \int_0 ^\infty \int_0 ^\infty {\rm d} S {\rm d} t \exp( - u {\cal S} - p t) P({\cal S}, t)$ be the double Laplace transform ${\cal S} \to u$ and $t \to p$ of the PDF $P({\cal S}, t)$. Then we derive a Montroll-Weiss \cite{MW} like
equation which reads
\begin{equation}
\hat{P} (u,p ) = { \hat{\Phi}(u,p) \over 1 - \hat{\phi}(u,p) } .
\label{eqMW}
\end{equation}
Here $\hat{\phi}(u,p)$ is the double Laplace transform of $\phi(s,\tilde{\tau})$
and similarly for the pair $\hat{\Phi}(u,p)$ and
\begin{equation}
\Phi (s, t _B) = \int_0 ^{v_{{\rm max}}} {{\rm d} v \over v_{{\rm max}}}
\exp \left( - R(v) t_B \right) \delta\left(s- t _B v^2\right).
\label{eqG06}
\end{equation}
This 
 term stems from the contribution
to the action $s=v^2 t_B$
 from the last increment in the sequence,  namely from the backward recurrence time $t_B$, while $\exp(- R(v) t_B)$ is the probability of not jolting in that 
time interval. 
We now
promote the use of the Montroll-Weiss like  tool Eq. (\ref{eqMW})  in the context of ergodicity, following the next
steps.

To investigate the ergodic properties of the process
 we define the dimensionless random
 variable
$\Upsilon = \overline{E}_k / \langle \overline{ E}_k\rangle$.
We first  focus on the case when the observable is integrable, namely $1/3<\gamma<1$. 
In standard ergodic theories, found here if $1<\gamma$, the distribution of $\Upsilon$ will approach a delta function centred on unity.
By defintion  $\Upsilon = {\cal S}(t) / ( t 
\langle \overline{ E}_k(t)\rangle)$, 
and here an analysis of Eq. (\ref{eqMW}) is useful, since it yields the distribution of ${\cal S}$ and then the distribution of the sought after $\Upsilon$. We find the universal behavior that
the PDF of $\Upsilon$ denoted $\mbox{ML}(\Upsilon)$ 
 is given by the Mittag-Leffler (ML) law 
\begin{equation}
\mbox{ML} \left( \Upsilon \right) = { [\Gamma \left( 1 + \gamma \right)]^{1/\gamma} \over \gamma \Upsilon^{1 + 1/\gamma} } 
l_{\gamma,1} \left( { [\Gamma (1 + {1 \over \gamma} )]^{1/\gamma}  \over \Upsilon^{1/\gamma}} \right).
\label{eqML}
\end{equation}
Here $l_{\gamma,1}(.)$ is the one sided L\'evy PDF. This law which replaces 
Birkhoff's ergodic theory, is a concrete manifestation of the 
Darling-Kac theorem \cite{Darling,Aaronson}. Our physical approach was to show,
 how this law
is related to the laser cooling parameters, i.e. to $\gamma$.
 This law is valid for any
observable of interest, provided that it is integrable, for example
we verified this numerically for the energy observable Fig. \ref{figsix}
but also for the indicator function. In the limit
$\gamma\to 1$  $\mbox{ML}(\Upsilon)$  reduces to a delta function, as expected.

 
 We now  address the case when the energy in a  non-integrable observable,
namely $0<\gamma<1/3$.
The calculation of the mean energy cannot be made with the infinite density,
 since the result will diverge. Instead, here the ensemble
average  kinetic energy is found using the scaling solution Eq. 
(\ref{eq03}).
This means that the energy is now sensitive to the inner part of the velocity packet
\begin{equation}
\left\langle E_{k}(t)\right\rangle \sim \frac{\sin (\pi \gamma )}{\sin (3\pi
\gamma )}\frac{1}{\Gamma (2-2\gamma )}c^{2\gamma }t^{-2\gamma }.
\label{eqEAEk}
\end{equation}
This behavior is universal in the sense that this result does not depend on
$v_{{\rm max}}$, unlike Eq. 
(\ref{eqMEAN}).
In this case, a simple time integration gives the relation between time and
ensemble average 
$\langle \overline{E_{k}(t)}\rangle \sim \langle
E_{k}(t)\rangle / (1-2\gamma)$ which clearly differs
from the generic behavior found for integrable observables 
Eq.
(\ref{eqave07})
.

To characterize the fluctuations of the time averages 
it is useful
to define the ergodicity breaking parameter \cite{PRLctrw} 
\begin{equation}
\mbox{EB} = { \langle \overline{E_k}^2 (t) \rangle - \langle \overline{E_k}(t)\rangle^2 \over
\langle \overline{E_k}(t) \rangle^2} = 
{ \langle {\cal S}^2(t)  \rangle - \langle {\cal S}(t)  \rangle^2 \over \langle {\cal S} (t) \rangle ^2 },
\label{eqFLUC10}
\end{equation}
which is zero in the long-time limit,  if the process is ergodic, namely for $\gamma>1$. In the phase where energy is integrable $\mbox{EB}=
[{ 2 \Gamma^2(1+\gamma) - \Gamma(1 + 2 \gamma) ]/ \Gamma(1 + 2 \gamma)}$,
for $1/3<\gamma<1$, which as mentioned is universal in the sense that
it goes beyond the observable of interest. For example this describes also
the fluctuations of the number of renewals $N(t)$.

 Returning to the case $\gamma<1/3$, we need to find the second
moment of the action, which is obtained in Laplace $p$ space
after differentiation of Eq. 
(\ref{eqMW})
\begin{equation}
\langle \hat{{\cal S}}^2 (p) \rangle=  {\partial^2 \over \partial u^2}
{ \hat{\Phi}(u,p) \over 1 - \hat{\phi}(u,p) }|_{u=0}.
\end{equation}
This expression is analysed in the small $p$ limit,
 and we find asymptotically for large times and  for $0<\gamma<1/3$
\begin{equation}
\mbox{EB} = 
{ 2 \Gamma^2(2 - 2 \gamma) \over \Gamma(3 - 4 \gamma) } \left[ {
\sin^2(3 \pi \gamma) (1 - 5 \gamma) \over \sin (\pi \gamma) \sin (5 \pi \gamma)} + 3\gamma \right] - 1.
\label{eqNI13}
\end{equation}
This expression clearly differs from the $\mbox{EB}$ parameter found in the Darling-Kac phase $1/3<\gamma<1$. 
When $\gamma \to 0$ we find $\mbox{EB}=4/5$. In this limit
the particle maintains a constant speed for (nearly) all the observation time
so 
$\overline{E}_k = v^2$ and
hence using Eq. 
(\ref{eq03}),
$\lim_{\gamma \to 0} \mbox{EB}= \langle v^4 \rangle - \langle v^2 \rangle^2 / \langle v^2 \rangle^2=4/5$. 
The $\mbox{EB}$ parameter versus $0<\gamma<1$
 is plotted  in Fig. \ref{figsix}${\bf f}$  and it exhibits a clear transition
when $\gamma=1/3$. Thus  switching from the case when
energy is integrable to non-integrable, manifests itself in non-trivial
fluctuations
of the time averages. 

  We have  investigated
 semi-analytically the distribution of the time averages
also in the non-integrable phase $\gamma<1/3$ namely $\alpha>3$.
Here, clearly the Mittag-Leffler law is not valid any more. 
In Fig. \ref{figsix}
we present some of the main results of this mathematically 
challenging domain. 
For example,  we find 
$\lim_{\gamma\to 0} P(\Upsilon)= \Upsilon^{-3/2}/ (2 \sqrt{3})$
 for $\Upsilon<3$, 
otherwise $P(\Upsilon)=0$.  This diverges for small $\Upsilon$
 and
also  has a non-analytical cutoff at $\Upsilon=3$.
This result can be explained, as we did for the $\mbox{EB}$ parameter, noting that the atom  maintains a constant speed for practically all the duration
of the measurement i.e. 
in this limit  $\Upsilon=v^2/\langle v^2 \rangle$ and the PDF of $v$ 
is the uniform $f(v)$. 
 For the experimentally relevant case
$\alpha=4$ $(\gamma=1/4)$, 
we find that the distribution of $\Upsilon>0$ is 
half-Gaussian, see Fig. \ref{figsix}({\bf c}).  Interestingly this case also  marks a transition: 
the PDF of $\Upsilon$ diverges at the origin for any $\alpha>4$ and vanishes there for $3<\alpha<4$.
This means that for $\alpha>4$
the most likely
time average is actually found for
 cases where it  is much smaller than the ensemble average. 
Finally, for $\alpha=6$, a case also considered relevant for experiments, we find the solution in terms of a Fox-H function \cite{Prudnikov}, 
 $P(\Upsilon) \simeq
\frac{C}{\Gamma (\frac{3}{4})}H_{1,1}^{1,0}\left(
C\Upsilon \left\vert 
\begin{array}{c}
(\frac{1}{3},\frac{2}{3}) \\ 
(-\frac{1}{4},1)%
\end{array}%
\right. \right)$
with $C=\frac{3}{4\Gamma (\frac{5}{3})}$,
 which perfectly matches the simulation presented in Fig. \ref{figsix}({\bf d}).

%

We have seen how the non-normalizable
 infinite invariant measure ${\cal I}(v)$  replaces
in some sense the Maxwellian velocity distribution for thermal gases,
 which is of course
also invariant but perfectly normalizable. 
We showed how the ergodic properties of the system exhibit
three phases controlled by the fluorescence rate parameter $\gamma$. 
The 
 divergence of the mean trapping time at $\gamma=1$  is a well known factor
in the change of ergodic properties of a vast number of
physical systems.
 We highlighted a second novel
transition, 
 associated with the changeover  of an observable from being
integrable
to non-integrable, with respect to the non-normalised state  ${\cal I}(v)$, 
found at $\gamma=1/3$ for the energy observable. 
Similar behaviors can be found for other observables and as the applications
of infinite ergodic theory expand,  either for stochastic or deterministic
systems, we expect this type of transition to be wide spread. 

%
%

\begin{acknowledgments}
{\bf Acknowledgments:}
The support of Israel Science Foundation's grant 1898/17 is acknowledged 
(EB). 
This work was supported by the JSPS KAKENHI Grant No 240 18K03468 (TA).
GR thanks Tony Albers for helpful discussions. 
\end{acknowledgments}

\newpage

\begin{widetext}

\section{Supplementary Material}

\subsection{The Master Equation}

We briefly discuss the derivation of the non-normalized quasi steady-state
${\cal I}(v)$,
Eq. $(2)$ in the  main text,
and the scaling solution $g(x)$, Eq. $(3)$, see also \cite{CT,Bertin}.
Let $\rho(v,t)$ be the distribution of the speed of the atom.
As mentioned in the text, after a recoil event, 
the parent PDF of $v$ is $f(v)$.
As is well known, this does not imply that the steady-state
velocity is given by $f(v)$ since the fluorescence  rate $R(v)$ depends on
$v$. Note that in simulations, in the text and following \cite{CT} $f(v)$ was uniform, however, a variation will not change dramatically the main conclusions
of our work.  
Following \cite{CT}  the transition from state $v$ to $v'$ is given by
$W(v\rightarrow v') = R(v) f(v')$. The Master equation 
then reads
\begin{equation}
{\partial {\rho}(v,t) \over \partial t}  = -{ \rho(v,t)  \over \tau(v)} + f(v) \int_{0} ^\infty { \rho(v',t) \over \tau (v')} {\rm d} v',
\label{eqSM01}
\end{equation}
where the first  term describes the losses and the second the gains. 
Here $\tau(v)=1/R(v)$
is the life time of an atom in  state $v$.  
A time independent solution of this equation reads
$\rho^{*} (v) = \mbox{Const} \tau(v) f(v)$. If this solution is normalizable, we can determine the constant, and then we get the usual steady-state of the system.
However, we consider the opposite situation,
when  $\tau(v) \sim c v^{-\alpha}$ for $v \to 0$ and $\alpha>1$ and, as mentioned in the text, in current experiments $\alpha=2,4,6$ are relevant.
In this model the life time  $\tau(v)$ becomes long when the velocity is small, and hence this describes trapping of atoms at small velocities, which leads
eventually to efficient cooling. 
To solve the Master equation in the long-time limit, we invoke two types of solutions, which we later match.
For long times and not too small $v$ we have 
\begin{equation}
\rho(v,t) \sim {b \tau(v) f(v) \over t^{1 - \xi}}.
\label{eqSM02}
\end{equation}
The numerator has the structure of the usual  steady-state and $b$ and $\xi$
are obtained  by matching. 
On the other hand for small $v$, of the order of $t^{-1/\alpha}$ we have a 
scaling solution $\rho(v,t) \sim t^{1/\alpha} g (v t^{1/\alpha})$. Inserting this
ansatz into the Master equation, we obtain an integral equation for $g(x)$.
Following this procedure and matching the two solutions
 one arrives at  Eqs. (2,3)
in the main text, 
which are the starting point of our work. We extensively verified
these solutions with numerical simulations.
Specifically one finds the exponent  $\xi=1/\alpha$ \cite{CT}.

\subsection{
Figure preparation}

To prepare plots, we  simulate the process, using an
initial condition for the speed $v$  drawn from a uniform distribution
in the interval $(0,1)$ so $v_{{\rm max}}=1$. 
The waiting time is a random variable following 
the exponential distribution with mean $\tau(v)$. In
 numerical simulations, a uniform random variable $x$ on $[0,1]$ is used  and we  transform $x$ into $y = - \tau(v) \ln x$ which gives the random time
between collision updates.
The speed of the atoms is fixed between these events. 
The next velocity and waiting time  are determined in the same way as the above.
Recall that $\tau(v)= c v^{-\alpha}$ and in simulations we use $c=1$, which does not alter the main conclusions of the paper.

We now provide further details of the figure. 
\begin{itemize}
\item{Fig. $1({\bf a})$.} PDF $P(\Upsilon)$
 of the normalized
 time averaged  energy $\Upsilon = \overline{E}_k (t) / \langle \overline{E}_k (t) \rangle$ 
is plotted for $\alpha=4/5$.  The measurement time is
 $t= 10^9$. In the limit of infinite
measurement time, this PDF will approach a delta function, since here we are considering the normal ergodic phase $\gamma>1$ (because $\gamma=1/\alpha=5/4$).
The number of trajectories used  was $10^4$. 
To obtain $\Upsilon$, we calculate  $Y(t)=S(t)/t$, where $S(t)$ is the action defined in the text.  This is performed  
 for each trajectory and the mean is obtained using
an average over the ensemble  of  trajectories.
$Y(t)$ divided by the mean gives the random variable
 $\Upsilon$. Repeating many times we find the histogram for $\Upsilon$. 

\item{
Fig. $1{\bf (b)}$.}
Same as the  previous plot, the  PDF $P(\Upsilon)$ of 
the  time averaged  energy 
 $v^2(t)$, but now  $\alpha=5/4$ and hence $\gamma=4/5$,
the measurement time is $t=10^6$ and the number of trajectories $10^6$.
Since $1/3<\gamma<1$ the theoretical PDF of $\Upsilon$ is the Mittag-Leffler distribution
with index $4/5$, see Eq. (12) in the Letter. As explained in the text,
one sided L\'evy stable functions $l_{\gamma,1}(x)$  are used to plot this
PDF. Specifically
 $l_{\gamma,1}(x)$ is by definition the inverse Laplace transform of 
$\exp(-u^\gamma)$, and  $x>0$.
 In turn these functions are implemented in programs like Mathematica, hence
 it is easy 
to plot the theoretical prediction for $P(\Upsilon)$. 
As shown in the figure, this perfectly matches
the simulations without fitting.  

\item{Fig. $1{\bf (c)}$.}
 PDF $P(\Upsilon)$
for $\alpha=2$ and $\alpha=4$.
The measurement time is fixed as $t=10^{11}$, the number of trajectories is
$10^6$. For $\alpha=2$ note that the Mittag-Leffler PDF, defined in the text, ishalf-Gaussian. While the calculation of the PDF of $\Upsilon$ for $\alpha=4$, which also gives
a half-Gaussian distribution, is based on the Mellin transform (see below). 
 As in other sub-plots theory and simulations
perfectly match. 

\item{Fig. $1{\bf (d)}$.} Same as the other plots however now $\alpha=6$,
$t=10^{15}$ and the number of realizations $10^6$.  The results
of  simulations, presented as a
histogram, perfectly match the theory given in the text. 
As we will soon explain, 
the latter is expressed in terms of a Fox H-function, which gives 
$P(\Upsilon) =
\frac{C}{\Gamma (\frac{3}{4})}H_{1,1}^{1,0}\left(
C\Upsilon \left\vert
\begin{array}{c}
(\frac{1}{3},\frac{2}{3}) \\
(-\frac{1}{4},1)%
\end{array}%
\right. \right)$
with $C=3/(4\Gamma (5/3))$.
After rewriting this expression in terms of a Meijer G-function
we can plot the result using Mathematica.

\item{Fig. $1{\bf (e)}$.} Same as the other plots,  now for simulations
with $\alpha=50$. This in turn is compared with the $\alpha\to \infty$ limit of our theory presented in the main text.

\item{Fig. $1{\bf (f)}$.} $\mbox{EB}$
 parameter for the energy observable $v^2(t)$. 
The process is the same as in other figures though now we vary $\gamma$,
namely we present $\mbox{EB}$ versus $\gamma$.
The $\mbox{EB}$ parameter is defined in Eq. (14) in the main text. 
The number of the trajectories used to produce the figure  is $10^6$
 except for $\gamma > 1$ 
 where it is $10^5$. The method of calculating
the time average is the same as in the previous plots. 
The theoretical predictions, plotted in the figure,
 are given in the text and they read 
\begin{equation}
\mbox{EB}= \left\{
\begin{array}{l l}
{ 2 \Gamma^2(2 - 2 \gamma) \over \Gamma(3 - 4 \gamma) } \left[ {
\sin^2(3 \pi \gamma) (1 - 5 \gamma) \over \sin (\pi \gamma) \sin (5 \pi \gamma)} + 3\gamma \right] - 1,  & \ \  0<\gamma<1/3 \\
 { 2 \Gamma^2(1+\gamma) \over  \Gamma(1 + 2 \gamma)} -1, & \ \  1/3< \gamma<1 \\
0,  & \  \ 1<\gamma
\end{array}
\right.
\end{equation}
where $\Gamma(.)$ is the Gamma function. 
Here we have three phases: energy is non-integrable with respect to the infinite measure
$0<\gamma<1/3$, the energy is integrable $1/3<\gamma<1$  (in the Letter this 
 was called also the Darling-Kac phase),
and finally $1<\gamma$ where standard ergodic theory holds and
the $\mbox{EB}$ parameter is equal zero. 

\end{itemize}

\section{Details on the derivation of $P(\Upsilon)$}

 We provide some details on the derivation of $P(\Upsilon)$ focusing on
the phase where the energy observable is non-integrable with respect to
the infinite measure. The calculations are lengthy, and hence here
we provide only an outline. We plan a longer publication
which will provide further technical details. 

The double Laplace transform of the action propagator $P(S,t)$ is given by $%
\widehat{P}(u,p)$ in Eq. (10) in the Letter. 
In the limit $t\rightarrow \infty $, and
equivalently $p\rightarrow 0$, these quantities take corresponding scaling
forms, which inserted into the Laplace transform relation yields%
\begin{equation}
\frac{1}{p}g_{\alpha }(\frac{u}{p^{1-2/\alpha }})=\int\limits_{0}^{\infty
}dS\int\limits_{0}^{\infty }dt\;e^{-uS}e^{-pt}\frac{1}{t^{1-2/\alpha }}%
f_{\alpha }(\frac{S}{t^{1-2/\alpha }}),  \label{double Laplace scaling}
\end{equation}%
where $f_{\alpha }(x)$ and $g_{\alpha }(y)$ are the scaling function in
action space and in the corresponding Laplace space, respectively.
Substituting $x=\frac{S}{t^{1-2/\alpha }}$ and setting $p=1$ turns Eq. (\ref%
{double Laplace scaling}) into the integral equation%
\begin{equation}
g_{\alpha }(y)=\int\limits_{0}^{\infty }dx\;K_{\alpha }(yx)f_{\alpha }(x)
\label{Fredholm}
\end{equation}%
relating $g_{\alpha }(y)$ and $f_{\alpha }(x)$ via this convolution
transform with kernel 
\begin{equation}
K_{\alpha }(x)=\int\limits_{0}^{\infty }dt\;e^{-xt^{1-2/\alpha }}e^{-t}.
\label{kernel}
\end{equation}%
For the scaling function $g_{\alpha }(y)$, with $\alpha >3$, i.e. in the
non-Darling-Kac phase, we obtain from the  $p\rightarrow 0$ limit of $%
\widehat{P}(u,p)$, Eq. (10) in the text, the following exact form 
\begin{equation}
g_{\alpha }(y)=\frac{\int\limits_{0}^{\infty }\frac{1}{1+yz^{2}+z^{\alpha }}%
dz}{\int\limits_{0}^{\infty }\frac{1+yz^{2}}{1+yz^{2}+z^{\alpha }}dz}.
\label{gy exact}
\end{equation}%
The goal is to obtain from this by inversion of Eq. (\ref{Fredholm}) the
scaling function $f_{\alpha }(x)$, which is a rescaled version of the
limiting probability density $P_{\alpha }(\Upsilon )$ of the normalized time
average $\Upsilon $ (see main text). Such an inversion can be achieved in
principle by a Mellin transform of both sides of Eq. (\ref{Fredholm})
resulting in [\cite{Polyanin}, p.997]
\begin{equation}
\widetilde{g}_{\alpha }(s)=\widetilde{K_{\alpha }}(s)\widetilde{f}_{\alpha
}(1-s),  \label{Mellin product}
\end{equation}
where $\widetilde{g}_{\alpha }(s)=M[g_{\alpha
}(y);s]=\int\limits_{0}^{\infty }dy\;g_{\alpha }(y)y^{s-1}$ is the Mellin
transform of $g_{\alpha }(y)$, and $\widetilde{K_{\alpha }}(s)$ and $%
\widetilde{f}_{\alpha }(s)$ is defined analogously. Solving Eq. (\ref{Mellin
product}) for $\widetilde{f}_{\alpha }(s)$ and applying the inverse Mellin
transformation gives 
\begin{equation}
f_{\alpha }(x)=M^{-1}[\widetilde{f}_{\alpha }(s);x]=M^{-1}[\frac{\widetilde{g%
}_{\alpha }(1-s)}{\widetilde{K}_{\alpha }(1-s)};x],  \label{inverse Mellin}
\end{equation}%
and finally by rescaling the desired limit density is obtained as%
\begin{equation}
P_{\alpha }(\Upsilon)=C_{\alpha }f_{\alpha }(x)|_{x=C_\alpha \Upsilon},  \label{rhofx}
\end{equation}%
where $C_{\alpha }=\left\langle x\right\rangle _{f_{\alpha
}}=\int\limits_{0}^{\infty }dx\;xf_{\alpha }(x)$. The rescaling implies
that the mean is $\left\langle \Upsilon \right\rangle _{P_{\alpha
}}=\int\limits_{0}^{\infty }d \Upsilon\;\Upsilon P_{\alpha }(\Upsilon)=1$, as requested. While in
principle along these steps the problem of calculating the limit
distribution $P_{\alpha }(\Upsilon)$ is solved, a fully analytic solution is
available only in two cases, $\alpha =4$ and $\alpha =\infty $. For $\alpha
=4$ the integrals in Eq. (\ref{gy exact}) can be evaluated by residue
calculus yielding after some calculations the simple result%
\begin{equation}
g_{4}(y)=\frac{1}{1+y},  \label{g4 final}
\end{equation}%
with Mellin transform $\widetilde{g}_{4}(s)=\Gamma (s)\Gamma (1-s)$. Since
the Mellin transform $\widetilde{K_{\alpha }}(s)$ of the integral kernel is
also known in full generality, $\widetilde{K}_{\alpha }(s)=\Gamma (s)\Gamma
(1-(1-2/\alpha )s)$, we get the quotient in Eq.(\ref{inverse Mellin}), and
in addition we can invert from Mellin space to obtain $f_{4}(x)=\frac{1}{%
\sqrt{\pi }}e^{-\frac{x^{2}}{4}}$. The scaling factor $C_{4}=\left\langle
x\right\rangle _{f_{4}}$ follows from the general relation between the $n$%
-th derivative $g_{\alpha }^{(n)}(y=0)$ and the moments $\left\langle
x^{n}\right\rangle _{f_{\alpha }}$ of $f_{\alpha }(x)$ 
\begin{equation}
g_{\alpha }^{(n)}(0)=K_{\alpha }^{(n)}(0)\left\langle x^{n}\right\rangle
_{f_{\alpha }}=(-1)^{n}\Gamma (1+n(1-\frac{2}{\alpha }))\left\langle
x^{n}\right\rangle _{f_{\alpha }},  \label{moments general}
\end{equation}%
which follows directly from Eq. (\ref{Fredholm}). For $\alpha =4$ we get $%
C_{4}=\left\langle x\right\rangle _{f_{4}}=\frac{2}{\sqrt{\pi }}$, so that
we obtain for $P_{4}(\Upsilon)$ according to Eq. (\ref{rhofx}) a half Gaussian
distribution
\begin{equation}
P_{4}(\Upsilon)=\frac{2}{\pi }e^{-\frac{\Upsilon^{2}}{\pi }}  \label{rho4}
\end{equation}
as exact result, which is displayed together with the results from numerical
simulations in Fig. 1 ${\bf c}$. 
We can proceed similarly for $\alpha \rightarrow
\infty $, because all integrals and Mellin transforms are known exactly also
in this case yielding  
\begin{equation}
g_{\infty }(y)=\frac{1}{\sqrt{y}}\arctan \sqrt{y}  \label{gyinf}
\end{equation}%
and eventually  
\begin{equation}
P_{\infty }(\Upsilon)=\frac{1}{2\sqrt{3}}\frac{1}{\sqrt{\Upsilon}}\theta (3-\Upsilon),
\label{rhoinf}
\end{equation}%
which is also plotted in Fig. 1${\bf e}$. 

For the experimentally also relevant case $\alpha =6$ we can still get from
Eq. (\ref{gy exact}) by residue calculus a fully analytic expression for $%
g_{6}(y)$, but the result is very lengthy involving 3rd roots etc., and it
does not simplify as in the case  $\alpha =4$. Therefore one cannot
analytically calculate its Mellin transform. This led us to find an
approximate scaling function $g_{6}^{\approx }(y)$, which deviates only
little from the true function $g_{6}(y)$, but for which the Mellin transform
is known analytically. The general ideas is to match in $g_{6}^{\approx }(y)$
the true small $y$-behavior and the large $y$-asymptotics of $g_{6}(y)$,
which we know analytically from an analysis of Eq. (\ref{gy exact}). The
simple form 
\begin{equation}
g_{6}^{\approx }(y)=(1+\frac{2}{3}y)^{-\frac{3}{4}}  \label{g6yapprox}
\end{equation}%
shares with the exact function $g_{6}(y)$ the identical first and second
derivative at $y=0$, and the exponent of the asymptotic behavior for $%
y\rightarrow \infty $. The relative deviation of  $g_{6}^{\approx }(y)$ from 
$g_{6}(y)$ is strictly less than 4.2\%. This function $g_{6}^{\approx }(y)$
can be Mellin transformed  \cite{Prudnikov}, and leads via  Eq. (\ref{inverse
Mellin}) and subsequent rescaling to an analytical expression for $%
P_{6}^{\approx }(\Upsilon)$, which can be expressed in terms of a Fox-H function as%
\begin{equation}
P_{6}^{\approx }(\Upsilon)=\frac{3}{4\Gamma (\frac{5}{3})\Gamma (\frac{3}{4})}%
H_{1,1}^{1,0}\left( \frac{3}{4\Gamma (\frac{5}{3})}\Upsilon\left\vert 
\begin{array}{c}
(\frac{1}{3},\frac{2}{3}) \\ 
(-\frac{1}{4},1)%
\end{array}%
\right. \right) .  \label{rho6xapproxfinal1}
\end{equation}%
As mentioned, after rewriting this expression in terms of a Meijer G-function
 [\cite{Prudnikov},
p.629], we can plot the result with Mathematica, as shown in Fig. 
1 ${\bf d}$,  with
no visible deviations from the result obtained by numerical simulations of
the process.

\end{widetext}

\end{document}